\documentstyle[epsfig, twocolumn]{esa_sp_latex}

\begin{document}
%

\parindent 0pt
\parskip 10pt plus 1pt minus 1pt
\hoffset=-1.5truecm
\topmargin=-1.0cm
\textwidth 17.1truecm \columnsep 1truecm \columnseprule 0pt 

\title{\bf PROPERTIES OF THE HARD X-RAY EMISSION FROM THE BLACK HOLE
CANDIDATES: CYGNUS X-1 AND 1E1740.7-2942}

\author{{\bf S.~Kuznetsov, M.~Gilfanov, E.~Churazov, R.~Sunyaev,
I.~Korel, N.~Khavenson,}\\ {\bf A.~Dyachkov, B.~Novikov}\vspace{2mm} \\
Space Research Institute, Russian Academy of Sciences,
Profsouznaya 84/32, Moscow 117810, Russia \vspace{2mm} \\
{\bf J.~Ballet, P.~Laurent, M.~Vargas, A.~Goldwurm} \vspace{2mm} \\
Service d'Astrophysique, DAPNIA/DSM, Bt 709, CEA Saclay,
91191 Gif-sur-Yvette Cedex, France \vspace{2mm} \\
{\bf J.P.~Roques, E.~Jourdain, L.~Bouchet, V.~Borrel} \vspace{2mm} \\
Centre d'Etude  Spatiale des Rayonnements (CNRS/UPS) 9,
avenue du Colonel Roche,\\ BP 4346, 31028 Toulouse Cedex, France}
\maketitle

\begin{abstract}

The entire dataset of the GRANAT/SIGMA observations of Cyg X-1 and
1E1740.7-2942 in 1990-1994 was analyzed in order to search for
correlations between primary observational characteristics of the hard X-ray
(40-200 keV) emission - hard X-ray luminosity $L_X$, hardness of the
spectrum (quantified in terms of the best-fit thermal bremsstrahlung
temperature $kT$) and the {\em rms} of short-term flux variations.
 
Although no strict point-to-point correlations were detected certain general
tendencies are evident. It was found that for Cyg X-1 the spectral hardness
is in general positively correlated with relative amplitude of short-term
variability. The correlation of similar kind was found for X-ray transient
GRO J0422+32 (X-ray Nova Persei 1992, \cite{fin}) and recently for GX339-4 
(\cite{trud}).
 
For both sources approximate correlation between $kT$ and $L_X$ was found.
At low hard X-ray luminosity - below $\sim 10^{37}$ erg/sec - the $kT$
increases with increase of $L_X$. At higher luminosity the spectral hardness
depends weaker or does not depend at all on the hard X-ray luminosity. The
low luminosity end of these approximate correlations (low $kT$ and low
$rms$) corresponds to extended episodes of very low hard X-ray flux occurred
for both sources during SIGMA observations. \vspace {5pt} \\


\end{abstract}

\section{INTRODUCTION}

Cygnus~X-1 was first dynamically proven black hole candidate in the Galaxy
and its X--ray properties were studied in a number of occasions during last
two decades.  The source exhibits flux variations on all time-scales from
years to milliseconds. Two distinct spectral states of the source were
identified (\cite{tan72}; \cite{holt76}; \cite{oga82}): the ``Low State''
(LS) and the ``High State'' (HS). During the LS emission from the source in
X--ray domain is dominated by the hard spectral component observed up to
several hundred keV. The HS spectrum is much softer and is dominated by the
soft spectral component.  Most of the time ($\sim 90\%$) the source was
found in the LS (e.g. \cite{liang84}).  Basing on the observations of HEAO-3
Ling et al. (1987) proposed to distinguish three sub-states of the LS -- the
so called $\gamma$--states characterized by different intensity and spectral
properties of the X- and $\gamma$-ray emission.

The source 1E1740.7-2942 located 50 arcmin apart from the Galactic Center is
the hardest X-ray source in this region. On the basis of its X-ray
properties it was suggested that this source too contains a black hole
(\cite{rs91a}). Its spectral shape and X-ray luminosity are quite similar to
that of Cygnus~X-1 in its ``nominal'' $\gamma_2$\ state (\cite{rs91b}).
 
We report on the results of search for the correlation between
primary observational characteristics of the hard X-ray emission during low
state for these two sources: hard X-ray luminosity, hardness of the spectrum
and the amplitude of short-term flux variations. The preliminary results of
this analysis were presented earlier by Kuznetsov et al. (1995) and Ballet
et al.  (1996). Similar results were also obtained for Cyg X-1 by Crary et
al.  (1996) basing on the BATSE data.

\section{THE DATA ANALYSIS}

Only observations performed under nominal background and instrument
conditions were used for the analysis. Since the timing analysis is more
sensitive to nonstandard background conditions, in the case when timing
analysis was impossible or ambiguous the dataset was excluded from the
spectral analysis too.

In order to quantify the hardness of the source emission in the 40-200 keV
energy range the optically-thin thermal bremsstrahlung model (\cite{kel})
was chosen. Although possibly having no direct physical relation to origin
of the hard X-ray emission in compact sources, it provides a good
approximation to the observed spectra and characterizes the spectral shape
by a single parameter -- the best-fit temperature. The relative error of the
observed spectra approximation by this model is less than 10\% in the 40-200
keV energy band.  The model was applied to the pulse-height spectra averaged
according to chosen way of the data binning as described in next two
sections.  The 40--200 keV luminosity was calculated using the best fit
model and assuming distance of 2.5 and 8.5 kpc for Cyg X-1 and
1E1740.7--2942 correspondingly.

\begin{figure}[hp]
\epsfig{file=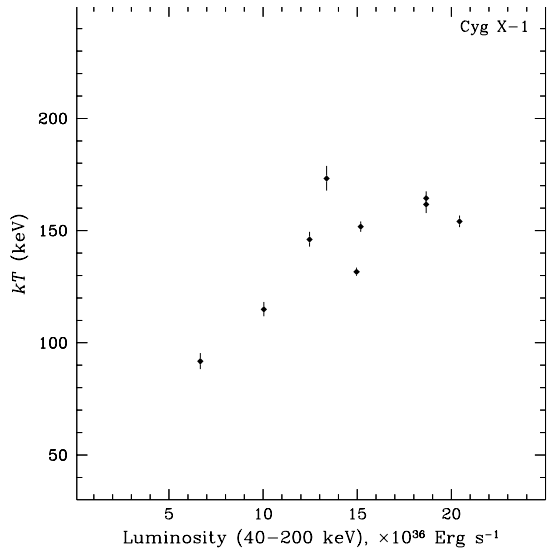, width=7.5cm,
 bbllx=20pt, bblly=160pt,bburx=570pt, bbury=695pt, clip=}
\caption{\em The best-fit bremsstrahlung temperature plotted against hard X-ray
luminosity (40-200 keV) for Cyg X-1. The data were averaged over 1 to 20
days of consequential observations.  }
\label{fgr1}
\end{figure}

For the timing analysis the 4 s resolution 40--150 keV data (count rate from
entire detector) were used. The power density spectrum was obtained for each
individual SI exposure using the standard timing analysis technique
(\cite{klis}) and then converted to the units of relative {\em rms} using
the source intensity averaged for duration of the same SI exposure.  The
values of relative {\em rms} were further averaged according to chosen way
of the data binning. The aperiodic variability of the source was quantified
by fractional {\it rms} of the flux variation in the 0.01-0.1 {\it Hz}
frequency band. This range corresponds to the flat part in the Cygnus~X-1
power density spectrum and represents its most variable part. The timing
analysis was not performed for 1E1740.7--2942 since this source is located
in the crowded Galactic Center region.

It should be noted that opposite to the spectral analysis which utilizes the
imaging capability of the SIGMA telescope the 4 s resolution data used for
the timing analysis does not possess spatial resolution and the count rate
from the entire detector was analyzed.  Numerous tests confirmed that under
standard background conditions possible contamination of the Cyg X-1 power
density spectrum by background events in the 0.01-0.1 {\it Hz} frequency
range can be neglected.

\section{RESULTS}

The light curves of both sources have a complex structure with short term
(time scales of days to weeks) variations superimposed on long term (time
scale of years) intensity changes of generally larger relative amplitude.
The SIGMA observations provided on one hand rather sparse time coverage --
especially for Cyg X-1, and, on the other, a limited time resolution
restricted by the instrument time resolution (several hours for spectral
information) and, especially for 1E1740.7--2942, by the accuracy of the
spectral and variability parameters estimation. The latter leads to
necessity of further grouping of the data.  In order to verify possible
effects of the data averaging two grouping methods were applied to the data
as described below.

\begin{figure}[hp]
\epsfig{file=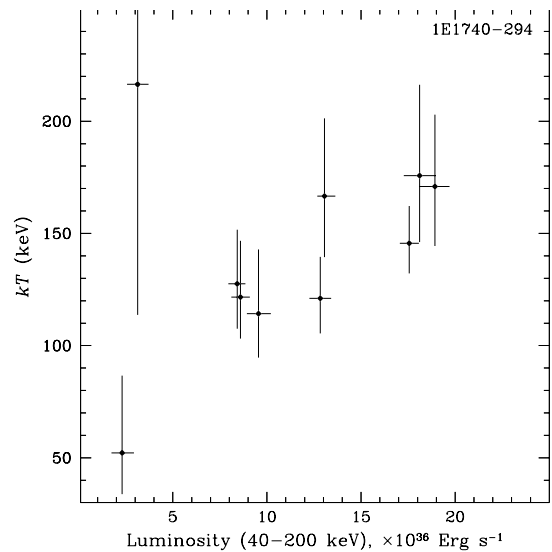, width=7.5cm,
 bbllx=20pt, bblly=160pt,bburx=570pt, bbury=695pt, clip=}
\caption{\em The best-fit bremsstrahlung temperature plotted against hard X-ray
luminosity (40-200 keV) for 1E1740.7--2942. The data were averaged over 10 to
60 days of consequential observations.  }
\label{fgr2}
\end{figure}

\subsection{Grouping by observational sets.}

In order to study the source behavior on the time-scale of months we have
averaged the data acquired during individual observational sets.  Typical
time span for each data point was $\sim$1 day to $\sim$1 month. Results are
shown in Fig. 1, 2 and 3.

\begin{figure}[hp]
\epsfig{file=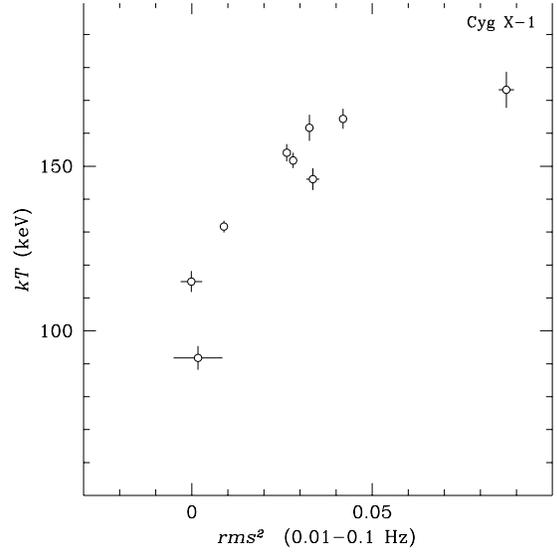, width=7.5cm,
 bbllx=20pt, bblly=160pt,bburx=570pt, bbury=694pt, clip=}
\caption{\em The best-fit bremsstrahlung temperature plotted
against $rms^{2}$\ of the short-term flux variations in the 0.01--0.1 Hz
frequency range for Cygnus~X-1. The data were averaged in the same way
as in Figure 1.
}
\label{fgr3}
\end{figure}

The relation between spectral hardness and luminosity is shown in Fig. 1
and 2. As it could be seen from Fig. 1 and 2 a trend is present in the
data for both sources -- the $kT$ generally increases as the hard X-ray
luminosity increases. It should be noted that for both sources the
correlation becomes less apparent if the data points corresponding to low
intensity episodes were removed -- it almost disappears for Cyg X-1 and
becomes nearly insignificant for 1E1740.7--2942.

The Fig. 3 shows a dependence of the best-fit bremsstrahlung temperature upon
the {\it rms} of the short-term flux variation for Cyg~X-1.  There is a
clear correlation between spectrum hardness and the amplitude of the
low-frequency noise (0.01-0.1 {\it Hz}) in the 40-150 keV energy band.

\subsection{Grouping according to intensity.}

The original data were regrouped according to the source intensity in the
following way. The entire range of the 40-200 keV flux variations was
divided into number of bins of the same width.  The mean energy spectrum and
power density spectrum corresponding to each intensity bin were calculated
by averaging over all individual datasets with intensity falling into the
given bin intensity range.

For Cyg X-1 16 intensity bins were chosen covering the 1.9 to 6.9 $10^{-2}$
cnt/sec/cm$^{2}$\ (0.5-1.8 Crab) intensity range. The regrouping procedure
was applied to the data of individual SI exposures (4-8 hours long - the
highest time resolution providing spectral information) each exposure being
treated as a separate dataset.  In the case of 1E1740-294, having 5 to 20
times lower signal to noise ratio the data averaged over each single
observation (comprised of 1-6 SI exposures with total duration of 4-34
hours) were treated as individual datasets to be regrouped. The intensity
range 0.3 to 5.6 $\times10^{-3}$ cnt/sec/cm$^{2}$\ (8-150 mCrab) was divided
into 10 intensity bins.

\begin{figure}[hp]
\epsfig{file=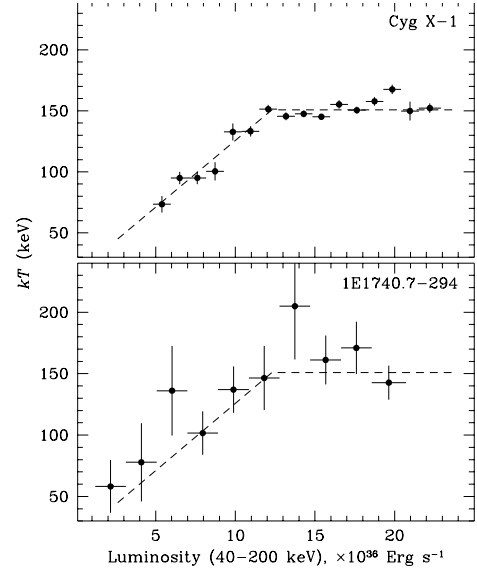, width=7.5cm,
 bbllx=20pt, bblly=150pt,bburx=570pt, bbury=701pt, clip=}
\caption{\em The best-fit bremsstrahlung temperature plotted against hard
X-ray luminosity (40-200 keV) for Cyg X-1 (upper panel) and 1E1740.7--2942
(lower panel).  The data were averaged according to the X-ray intensity as
described in the section 3.2.  The broken constant best-fit to Cygnus~X-1
data is shown by dashed line in both panels.  }
\label{fgr4}
\end{figure}

The dependence of the best-fit bremsstrahlung temperature upon the 40-200
keV luminosity for both sources is shown in Fig. 4. The similarity in the
behaviors of the two sources is apparent. Approximation of Cyg X-1 data by a
broken constant is shown by a dashed line. The same curve describes quite
well the data for 1E1740-294. The ``breaks'' in Fig. 4 correspond to the
same value of the luminosity.

It should be noted, that shown in Figure 3 are parameters derived from
averaged spectra for each intensity bin and the error bars are statistical
only. The analysis of individual datasets, corresponding to the given
intensity bin, performed with the more statistically significant Cyg X-1
data, revealed considerable dispersion of the best-fit parameters above the
level of statistical fluctuations. This dispersion is of the order of
$\sim$15\% of the values of the best-fit temperature shown in Figure 3.  It
is especially large in the $(0.8-1.2)\times 10^{37}$ erg/sec luminosity
range. Therefore, dependence, shown in Fig. 3 are not point-to-point
correlations, but rather represent some averaged pattern of the behavior of
the parameters.

\begin{figure}[hp]
\epsfig{file=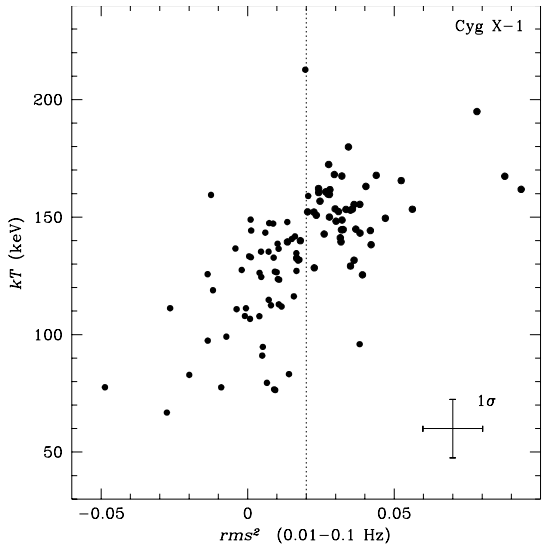, width=7.5cm,
 bbllx=20pt, bblly=160pt,bburx=570pt, bbury=695pt, clip=}
\caption{\em The best fit bremsstrahlung temperature plotted
against the $rms^{2}$ of the flux (40--150 keV) variation for Cyg~X-1. Each
data point correspond to individual exposure of $\sim$\ 4-8 hours long.
Average error is shown (actually the errors increase from the right to the
left). The $rms$\ level used for data classification (Figure 6) on the
basis of flux variability is shown by dotted line. 
}
\label{fgr5}
\end{figure}

\begin{figure}[hp]
\epsfig{file=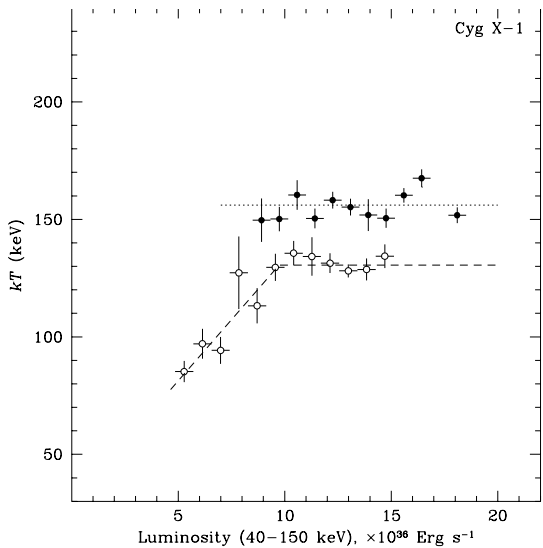, width=7.5cm,
 bbllx=20pt, bblly=160pt,bburx=570pt, bbury=695pt, clip=}
\caption{\em The hardness of the spectrum
versus hard X-ray luminosity (40-150 keV) for Cyg~X-1. The data were
averaged according to the X-ray luminosity. Open circles correspond to the
data with low noise level (points from the left half of Figure 5). Filled
circles correspond to the data with high noise level (Figure 5, the data
plotted right to dotted line). The broken constant best fit to low noise data
is shown by dashed line. The best fit to high noise level data is shown by
dotted line.  }
\label{fgr6}
\end{figure}

In order to illustrate the complexity of the source behavior we divided the
Cyg X-1 data (4-8 hours data segments) into two parts according to the level
of short term fluctuations (Fig. 5). The grouping procedure described above
(according to the source intensity) was then applied to the ``low rms'' and
the ``high rms'' data separately. The resulted relation between spectral
hardness and the hard X-ray luminosity is shown in Fig. 6. As it is seen from
Fig. 6 the $kT$ -- $L_X$ dependence splits into two distinct branches overlapping in
the luminosity.


\section*{ACKNOWLEDGMENTS}

The IKI co-authors would like to acknowledge partial support of this work by
INTAS grant 93-3364 and RBRF grant 96-02-18588-A. S.Kuznetsov was also
partially supported by the ISSEP grant S96-207.

\end{document}